# COMPLEX SYSTEM DESIGN WITH DESIGN LANGUAGES: METHOD, APPLICATIONS AND DESIGN PRINCIPLES

**Samuel Vogel[1] and Stephan Rudolph[2]**

[1] Institute for Aircraft Design, University of Stuttgart, Pfaffenwaldring 31, D-70569 Stuttgart
samuel.peter.vogel@gmail.com
+49 170 7078345

[2] Institute for Aircraft Design, University of Stuttgart, Pfaffenwaldring 31, D-70569 Stuttgart
stephan.rudolph@ifb.uni-stuttgart.de

**Abstract**
Graph-based design languages in UML (Unified Modeling Language) are presented as a method to encode and automate the complete design process and the final optimization of the product or complex system. A design language consists of a vocabulary (i.e. the digital building blocks) and a set of rules (i.e. the digital composition knowledge) along with an executable sequence of the rules (i.e. the digital encoding of the design process). The rule-based mechanism instantiates a central and consistent global product data structure (the so-called design graph). Upon the generation of the abstract central model, the domain-specific engineering models are automatically generated, remotely executed and their results are fed-back into the central design model for subsequent design decisions or optimizations. The design languages are manually modeled and automatically executed in a so-called design compiler. Up to now, a variety of product designs in the areas of aerospace (satellites, aircraft), automotive (space frame structures, automotive cockpits), machinery (robots, digital factory) and consumer products (coffeemakers) have been successfully accelerated and automated using graph-based design languages.
Different design strategies and mechanisms have been identified and applied in the automation of the design processes. Approaches ranging from the automated and declarative processing of constraints, through fractal nested design patterns, to mathematical dimension-based derivation of the sequence of design actions are used. The existing knowledge for a design determines the global design strategy (i.e. top-down vs. bottom-up). Similarity-mechanics in the form of dimensionless invariants are used for evaluation to downsize the solution for an overall complexity reduction. Design patterns, design paradigms (i.e. form follows function, or function follows form) and design strategies (divide and conquer) from information science are heavily used to structure, manage and handle complexity.

## 1 Introduction

The digitization of industrial processes, e.g. in the context of Industry 4.0, makes new design processes possible. The automation of the product development process promises a considerable increase in efficiency. Especially designs and decisions of the very early concept phase have a very large influence on the later life cycle costs of the product [1]. The development of modern and more competitive products requires to go even closer to the limits of what is physically feasible in order, for example, to squeeze the last bit of weight advantage or efficiency out of a product or system. Modern products are integrating typically multiple physical domains (mechanics, thermodynamic, electronics, logistics, …) as well as a lot of system levels consisting of sub-systems or parts that mutually build on each other. The combination of both, multiple domains together with a number of system entities results in a high level of design and process complexity that has to be handled. Digitized design processes can be used to cope with this complexity and to find more optimal product designs in even earlier project phases. This digitization mainly comprises the computer-aided synthesis of designs (CAD) including the automated generation of functional validation calculations and simulations (structural mechanics, fluid mechanics, controls...). In fact, a vir-tual product design shall be automatically generated and optimized based on given product requirements to optimally meet the performance targets. In this paper graph-based design languages are presented as a method to implement such digital and re-executable representations of (conceptual) design actions. At first, the method of graph-based design languages itself is explained. The method is proven for more than fifteen years and has been mainly developed in the Similarity Mechanics Group of the Institute for Statics and Dynamics (ISD), which moved now to the Institute of Aircraft Design (IFB) at the University of Stuttgart. Second, scientific applications are shown as well as an early industrial stage application. Finally, a collection of design principles to handle the complexity in product design is presented that have been identified in the scientific work with design languages over the past years.

## 2 Method

The method of graph-based design languages [2] is a further evolution step of generative, computer-based synthesis methods [3]. These synthesis methods can be divided in to string-based, shape-based and graph-based representations. From the viewpoint of the authors graph-based design languages belong to the most generic and abstract means of knowledge representation across different domains due to its graph representation. Alternative computer-based synthesis methods such as shape grammars [3] define a rule set on elementary shapes (vocabulary) which is recursively called in a production system to generate more complex shapes. Graph-based design languages expand this concept by generalizing the vocabulary to conceptual objects together with an adaptive procedural rule sequence.

**Philosophical Motivation** Rudolph gives a philosophical motivation for the design language concept in [4]. First it is observed that during the product design process different areas of concept, each with different levels of knowledge, are traversed. It is distinguished between the first area called 'believe' which covers uncertain design targets as simplicity, aesthetics or adequacy. The second concept is 'ability' which covers more concrete but still not exact formulated design aspects as Design for Manufacturing, Design for Assembly, Design for Recycling. The third concept covers exact aspects and is called 'knowledge'. It contains physical formulas and other reproducible, mathematically formalizable and provable laws and know-how. All three aspects have to be represented by one unified description to come to a proper formalization of the design process. Rudolph proposes a language-based representation that is closely related to natural languages which are a convenient candidate as they are able to cover all three conceptual areas presented above.

**Class Diagram (Vocabulary)** Instead of words in natural languages, parametrized objects that are instantiated from classes form the vocabulary of graph-based design languages. The graph-based design languages use the Unified Modeling Language (UML) as representation and modeling language whose roots are in the model-based software engineering [5]. The vocabulary is modeled as class diagram as shown in figure 2 top. The entities and parts that take part in the product design are casted into classes and enriched with parameters that represent e.g. physical or cost variables. Equations and constraints can be additionally modeled in the classes and are processed in an integrated solution path generator [6]. The equation and constraint network that is built on the instantiation of the classes is automatically solved in the solution path generator with an integrated computer algebra system.

**Rules (Grammar)** The engineering entities of the class diagram are rule-based instantiated into objects with specific parameter values. Graphical rules with a left-hand side (LHS) and a right-hand side (RHS) define the instantiation as manipulation on the instance diagram. The instance diagram contains the so-called design graph which is the unique abstract central-model of the product design. The design graph of an exhaust aftertreatment system is shown in figure 2 bottom. The design instances in the design graph are linked with each other according to the associations that are defined in the class diagram. The instance on the LHS of the graphical rule is looked-up in the design graph and replaced with the instance pattern on the RHS. The first rule is called 'axiom' and has an empty LHS as the design graph is empty in the beginning (figure 2 center left). In the 'axiom' the requirements and given boundary conditions are typically added. The graphical rule in figure 2 center right shows the incremental design step that adds a 'SCR-System' entity to the initial 'combustion engine'.

**Production System** The design process itself is then split into an incremental rule sequence. This predefined rule sequence is graphically modeled in an activity diagram and is called production system. Beside the graphical rules as building blocks an activity diagram can hierarchically host sub-activities which are activity diagrams themselves. As third object interface calls can be modeled that trigger the execution of engineering applications in so-called process chains that are described in the following section. The activity diagram in figure 2 top center shows from the left two calls of graphical rules followed by a two times alternating sub-activity and interface call. As fourth element which is not shown so-called decision nodes are available to branch the rule sequences in the production system in dependence of the state of the design graph.

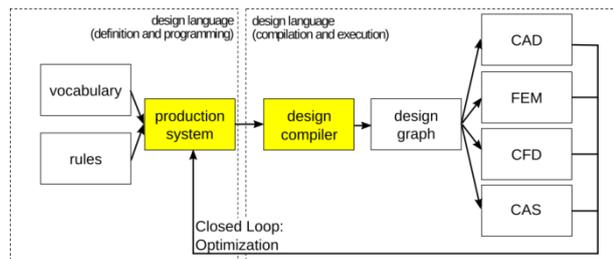

Figure 1 – Information architecture of graph-based design languages.

**Information Architecture** The figure 1 shows the information architecture that hosts the previous presented elements and shows their interaction. The class diagram (vocabulary), the rules as well as the production system represent, as digital blueprint of the design process, the design knowledge. The manual modeling of the class diagram, the rules and the production system is done in a so-called design compiler. For the presented work the commercially available Design Compiler 43V2 (www.iils.de) was used. The design language is executed in the design compiler which runs the production system. The design compiler provides engineering plugins that provide own class diagrams that can be used to model generic engineering tasks as CAD model creation and simulation model execution for evaluation and validation. These externally called engineering task called 'process chains' and are shown in figure 1 on the right side. The process chains extract the required information from the central design graph and create the CAD and simulation models automatically by executing model-to-model transformations. Results of the simulations runs are fed back to the design graph for sub sequent rules and operations.

## 3 Applications

A broad range of products has been designed using graph-based design languages. From applications in aerospace [7, 8] through consumer products [9] and off-road machinery components [10] up to automotive [11], the method has been successfully applied mainly in a scientific context. The scope of design languages has been prototypically extended up to downstream stages of the life-cycle by generating and designing the digital factory for a product in addition to the product itself [12]. Additional work has been done in the implantation of algorithms to automate engineering tasks as routing of cables and wires [8] and the automated creation of pipework in given installation spaces [13]. These intelligent wiring and piping algorithms become necessary as graph-based design languages fully automate the design process and therefore need to be able to create an intelligent integration and interaction of system components in given installation spaces for different product architectures.

### Class Diagramm

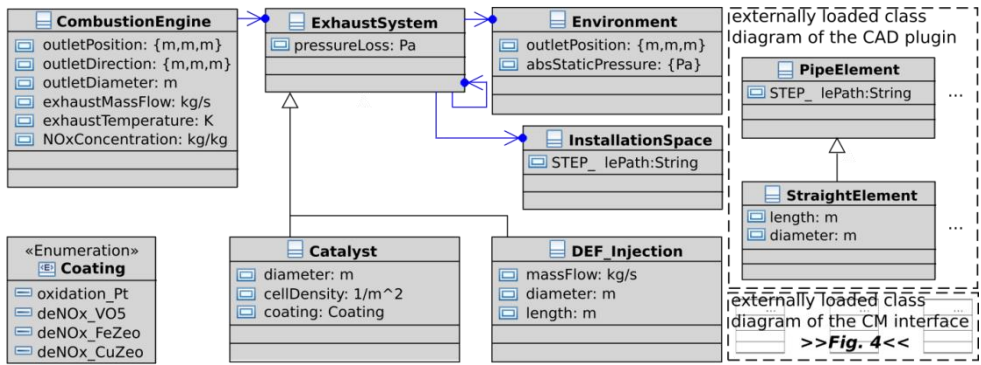

### Production System

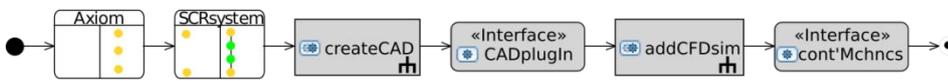

### Graphical Rules (if-then scheme)

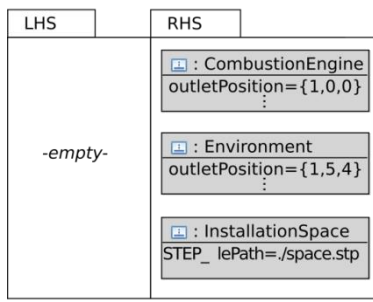
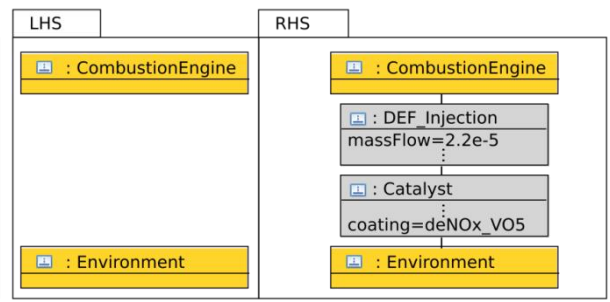

### Design Graph = Central Data Model (UML instance diagram)

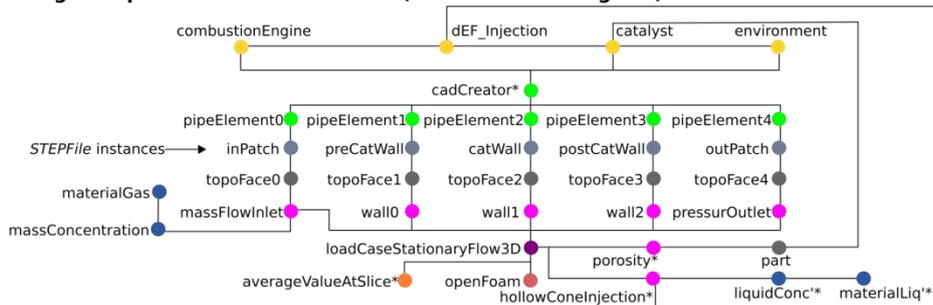

Figure 2 - Schematic graph-based grammar of an exhaust system and its main building blocks.

Figure 3 shows results of a graph-based design language that automates the layout design of an airplane's air cabin [8]. Beginning with the requirements, the designer can (manually) define the seating requirements based on a number of ratios. A ratio might define how many passengers share the same lavatory in a certain class. Together with the air planes main dimensions and the design language uses this as input conditions for a subsequent automated dimensioning and design of the air cabin (figure 3 bottom left). Beside the creation and integration of the 3D model (figure 3 top left) of the air cabin seating and infrastructure components (overhead bins, lavatories, kitchen,…) the systems and control units of the electrical infrastructure are automatically positioned for each seating arrangement (figure 3 top mid-left). Finally the electrical connections are automatically routed in the available and previously created 3D space according to additionally given constraints from e.g. electromagnetic compatibility.

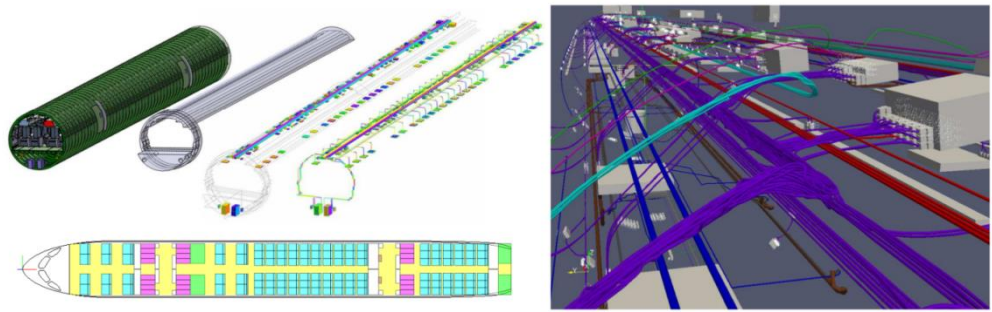

Figure 3 - Air cabin design with automated generation of electric and climate system. Reproduced from [8].

Using the graph-based design language reduces the time needed for an air cabin layout from many weeks and months to a few hours. The complexity of the interaction of the coupled systems and components (figure 3 right) is handled through the interplay of the production system and the design rules together with the intelligent algorithms to solve the integration tasks of positioning and wiring the electrical components. Different air cabin designs can thus be automatically evaluated and optimized in terms of total weight, cable length and wiring compatibility and validity [8].

Figure 4 shows the result and intermediate steps of a graph-based design language for creating a small satellite [7]. The design language starts from a given mission that defines payload, target orbit, energy and information demand of the payload as requirements. The rule set creates a complete virtual mockup based on the given requirements (figure 4 top row). The design language includes the automated creation and dimensioning of the control systems as well as the thermal validation as shown in figure 4 bottom left. Critical figures, such as the mass, energy and momentum balance, are calculated and balanced. A mission related communication system is chosen based on the mission figures. All critical subsystems are selected and in the subsequent integration step spatially arranged (packaged) and finally, with the previously mentioned routing algorithm, connected by wire (figure 4 bottom right). The integration of the components is validated for the defined orbits in terms of thermal loads arising from the incoming sun light as well as heat sources from components and systems (figure 4 bottom left).

Finally, figure 5 shows results of a design language for the automated functional optimization of SCR (selective catalytic reaction) exhaust aftertreatment systems, that reduce the emissions of internal combustion engines [10,13]. At the beginning a combustion engine is given in terms of mass flow and exhaust temperature as well as emissions and emission target. An installation space is also given, where the exhaust aftertreatment has to fit in (blue box figure 5 top right). The catalyst is analytically dimensioned in the beginning. Based on this size a CAD model of the catalyst housing is created. The catalyst box is automatically positioned in the installation space and finally a constraint pipework, based on standardized pipe bends, is created to connect the exhaust sys-tem components with the engine and the environment. The final CAD model is automatically meshed and a fluid simulation is created to determine the emission reduction efficiency as well as the pressure loss of the system (figure 5 center bottom). A finite element simulation to evaluate the thermal expansion of the system is created and executed as well (figure 5 center top). The whole design language is integrated into a design of experiments/optimization framework to conduct a design space exploration and to determine the pareto-front as best possible trade-off between competing design targets (pressure loss and emission reduction efficiency). The system was proven to scale very well on high performance clusters. Thousands of variants could be created and evaluated within a few weeks. Using this approach the physical limits of the exhaust aftertreatment architecture could be determine for the given requirements which is a very useful and valuable information, especially in an early project phase.

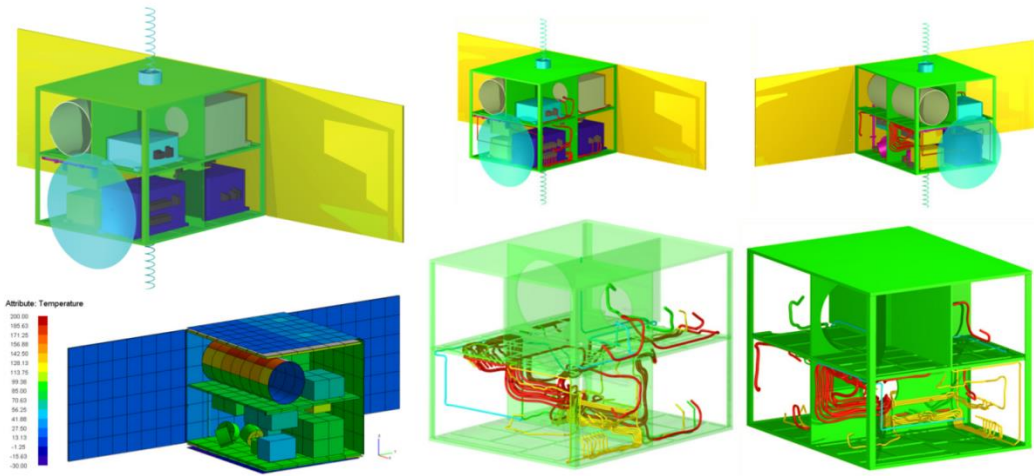

Figure 4 - Satellite design with graph-based design languages. Reproduced from [7].

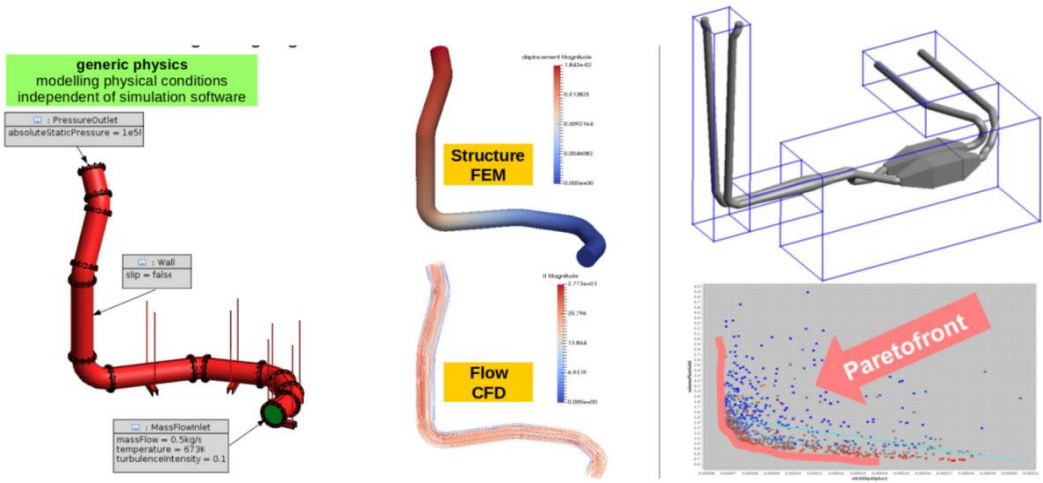

Figure 5 - Graph-based desing language for developing SCR exhaust aftertreatment systems in given installation spaces. Reproduced from [10, 13].

## 4 Design Principles

Different design strategies and mechanisms have been identified and applied in the automation of the design processes. Some principles have been identified by working on the explicit scientific problem of handling the complexity of engineering design and working on an engineering design theory. Others just emerged in application-centered design language projects as by-product.

**Top-Down and Bottom-Up** The work in [14] proposes to distinguish between top-down and bottom-up designs. The definitions in [14] can be directly quoted: "In the bottom-up approach it is attempted to achieve a higher level functionality by systematically combining basic building blocks into assemblies." A bottom-up design occurs especially in domains where only limited knowledge is available in advance. This phenomena often occurs in designs with non-linear physics, as fluid flows in the exhaust system example presented above, where small changes in geometry can produce significant changes in the overall physical behavior (e.g. flow separation). "In the top-down approach design synthesis ... is beginning with the requirements definition, the evolution of the design object is constituted out of subsequent decompositions from abstract conceptual descriptions into more detailed functional representations which finally find an embodiment into material components." The presented satellite example is an example for a top-down design where a priori knowledge is used to synthesize designs directly from the requirements.

**Dimensionless Evaluation** Rudolph identifies the evaluation of engineering objects as crucial challenge in the (automated) design of products [15] as the a chosen evaluation directly influences and determines the evolution of a product in optimization. Using the Pi-theorem with its dimensionless figures addresses the main problems of evaluation ("How can the evaluation of parts and components be found and represented? How are are partial results aggregated into a single evaluation? How are the goal criteria structured, arranged and are they complete? [15]). In this sense a complete description of a product entity in terms of design parameters and evaluation figures forms a valid evaluation: "any minimal description in the sense of the Pi-theorem is an evaluation" [15].

**Declarative Constraint Processing** In the process of designing complex products, consisting of many components, systems and spanning over multiple physical domains, large equation systems arise from the aggregation of analytical models that are part of each subsystem and component. This equation system is automatically assembled in the graph-based design language and subsequent automatically processed and solved in a so-called solution path generator [16]. Using the solution path generator a declarative processing of the equation and constraint system is ensured and the designer is freed from pre-defining a solution sequence. In addition, sensitivity analysis can be automatically executed to determine the critical design drivers and main dependencies within a product design [6].

**Self-Similar, Nested Design Patterns** The production system and a generic rule sequence was examined and a self-similar design pattern is postulated in [17]. It is called integrated design pattern and explained as: "The design cycle is an iterative sequence of design synthesis, analysis and evaluation. The design synthesis itself is subdivided into the definition of requirements, the decomposition, the formulation of functional solutions and the combinatorial exploration process." It combines aspects of both, the top-down and bottom up approaches into a "real world problem" pattern. It is observed that this pattern typically occurs in a hierarchical, nested manner on different granularity levels.

**Dimension-Based Design Sequence** The paper [10] deals with the problem of splitting the design process into a sequential step-by-step process. It is shown, that in order to find a global design optimum a dimensioning and integration of sub-systems and components has to occur simultaneously in one design step. But this implies that the optimal design parameter values have to be determined at once, forming a very high-dimensional optimization problem which practically can't be solved. A design sequence can then be deduced based on the mathematical dimension of the design parameters: When two systems are sequentially integrated, the (common) design pa-rameters of the lower dimensional system have to be fixed before fixing the design parameters of the higher dimensional system ("Begin with the system that has less degrees of freedom").

**Design Patterns and Paradigms** In the creation of a graph-based design languages it is possible to adopt design patterns [18] from object-oriented software engineering [18]. The use of object-oriented design patterns in design languages is made possible by the extension of classes by methods and interfaces proposed in [19]. The design patterns are solutions for recurring kind of problems in programming as for object creation, object composition (structural) and for object interaction (communication). They are typically made of two or more classes that are associated with each other and/or inherit behavior from each other, together with abstract methods in the classes and a defined interplay and usage of these methods to solve a specific problem. The 'Design for X' paradigms [4], mentioned above in the philosophical motivation, can also be seen as a kind of design pattern. As these paradigms are less specific than the previously presented design patterns from object-oriented software engineering they are part of the 'ability' concept area. This paradigms help the designer to structure the design process, at least mentally, when designing a graph-based design language. Nevertheless, it is often possible to transform the less specific paradigms in to rule-based heuristics when implementing a design language.

## 5 Discussion

In this paper an overview on graph-based design languages that automate recurring design tasks was given. The concept of graph-based design languages allows a graphical programming of design processes which helps to manage the complexity in product design by separating the data model from the operational procedural aspects in an object-oriented way. Using the object-oriented UML as modeling language allows a compact representation of the product entities that avoids redundancy through inheritance between and decomposition into classes. Storing engineering knowledge in incremental rules and adaptive rule sequences allows a hierarchical decomposition of the engineering process itself into smaller chunks that can be more easily captured and overviewed by the engineer even for complex products and systems. The presented example applications show that the method of graph-based design languages is able to solve real world engineering problems in a fraction of the time that would be necessary in manual engineering. In fact, the design time collapses to the addition of the (potentially concurrent) run-times of the algorithms which is close to the lower theoretical limit. Futhermore, a graph-based design language, when embedded in an optimization framework, is often able to find more optimal engineering solutions as can be found in a conventional engineering process. The presented language is able to capture engineering knowledge digitally which leads to a highly scalable and re-executable digital blueprint of recurring design tasks. Nevertheless, a significant upfront invest is necessary as implementing a product's generic design process is apparently more expensive than creating a few product designs by hand. The wide acceptance of this kind of modeling method in industry is at the moment therefore still often hindered by the traditional 'silo mentality' present in today's companies which look on the profitability of the individual business unit instead on the profitability of the company overall. The design principles are providing solutions to handle the occurring complexity in product design. They can be used on different levels and stages during the implementation of engineering processes. The presented principles are derived from frequently occurring patterns that can be reused and applied in different contexts. Handling the complexity in product design is crucial task as the complexity of the products themselves is increasing and even more domains and disciplines have to be considered to get a holistic evaluation, validation and optimization of a product's life cycle from cradle to grave. Graph-based languages are a method to push the limits of the still controllable complexity in industrial engineering further away.